\def\GlobalTapPreprint{1}
\documentclass{article}
\usepackage[T1]{fontenc}
\usepackage[utf8]{inputenc}
\usepackage{ismir}
\usepackage{amsmath,amssymb,cite,url,graphicx}
\usepackage{booktabs}
\usepackage{color}
\ifdefined\GlobalTapPreprint
  \usepackage{float}
\fi

\usepackage[bookmarks=false]{hyperref}
\hypersetup{colorlinks=true, urlcolor=blue, linkcolor=black,
  citecolor=black, filecolor=black}
\usepackage{xurl}

\newcommand{\RepoURL}{https://github.com/williambotticelli-wells/GlobalTap}
\newcommand{\SupplURL}{https://williambotticelli-wells.github.io/GlobalTap/supplementary.html}
\newcommand{\SourcesURL}{\RepoURL/blob/main/data/SOURCES.md}

\newcommand{\suppl}{\href{\SupplURL}{supplementary HTML}}
\newcommand{\suppldemos}{\href{\SupplURL\#audio-demos}{supplementary HTML}}

\ifdefined\GlobalTapPreprint
  \newcommand{\ApxRef}[2]{\begingroup\hypersetup{linkcolor=blue}%
    \hyperref[#2]{#1~\ref*{#2}}\endgroup}
  \newcommand{\MixedCorporaDetails}{listed in \ApxRef{Appendix}{app:mixed-source}}
  \newcommand{\RobustnessDetails}{Robustness analyses in \ApxRef{Appendix}{app:convergence}}
  \newcommand{\DemographicsDetails}{\ApxRef{Appendix}{app:demographics}}
  \newcommand{\BeatMetricsDetails}{\ApxRef{Table}{tab:app-beat-metrics}}
  \newcommand{\MetricalDetails}{full table in \ApxRef{Appendix}{app:metrical}}
  \newcommand{\ParticipantsSuppl}{\ApxRef{Appendix}{app:demographics}}
\else
  \newcommand{\MixedCorporaDetails}{\href{\SupplURL\#mixed-table}{listed in the supplementary material}}
  \newcommand{\RobustnessDetails}{\href{\SupplURL\#robustness}{Supplementary robustness analyses}}
  \newcommand{\DemographicsDetails}{the \href{\SupplURL\#demographics}{Supplemental participant demographics}}
  \newcommand{\BeatMetricsDetails}{the \href{\SupplURL\#beat-metrics}{Supplementary Table}}
  \newcommand{\MetricalDetails}{\href{\SupplURL\#metrical-audit}{full table in the supplementary material}}
  \newcommand{\ParticipantsSuppl}{the \suppl{} material}
\fi

\graphicspath{{figures/}{./figures/}}

\newcommand{\OverviewSchematicWidth}{0.82\textwidth}
\newcommand{\OverviewMapWidth}{0.78\textwidth}
\newcommand{\CountryPanelWidth}{0.9\columnwidth}

\newcommand{\symbolfootnote}[2]{%
  \begingroup
  \renewcommand{\thefootnote}{\fnsymbol{footnote}}%
  \footnotetext[#1]{#2}%
  \endgroup
}

\makeatletter
\renewcommand\section{%
  \@ismirsectiontrue
  \@startsection{section}{1}{\z@}%
    {-2.0ex \@plus -0.6ex \@minus -0.2ex}%
    {3pt \@plus 0.2ex}%
    {\large\bf\centering}}
\renewcommand\subsection{%
  \@ismirsectionfalse
  \@startsection{subsection}{2}{\z@}%
    {-1.5ex \@plus -0.6ex \@minus -0.2ex}%
    {3pt \@plus 0.2ex}%
    {\normalsize\bf\raggedright}}
\makeatother

\title{GlobalTap: Globally Diverse Beat Tracking Benchmark via Crowdsourced Tapping}

\def\authorname{W. T. Botticelli-Wells, M. Anglada-Tort, E. Celen,
R. Far, H. Lee, P. M. C. Harrison, and N. Jacoby}

\multauthor
  {William T. Botticelli-Wells$^{1}$ \quad
   Manuel Anglada-Tort$^{2}$ \quad
   Elif Celen$^{3}$ \quad
   Rena Far$^{4}$}
  {{\bf Harin Lee$^{5}$ \quad
   Peter M.~C. Harrison$^{5,\dagger}$ \quad
   Nori Jacoby$^{1,\dagger}$}\\
   $^{1}$ Cornell University, USA\\
   $^{2}$ Goldsmiths, University of London, UK\\
   $^{3}$ Max Planck Institute for Empirical Aesthetics, Germany\\
   $^{4}$ University of Toronto, Canada\\
   $^{5}$ University of Cambridge, UK\\
   {\tt\small ww577@cornell.edu}
  }

\begin{document}

\maketitle
\symbolfootnote{2}{These authors contributed equally and share senior authorship.}

\begin{abstract}
Existing beat-tracking benchmarks draw from a narrow range of musical
cultures, predominantly Western, and are difficult to scale due to
reliance on expert annotation. We address both limits with GlobalTap, a benchmark of 200
popular-music excerpts spanning 59 countries, annotated by 746 participants recruited across ten countries. The benchmark is created using a scalable online tapping pipeline that records
taps via audio for high temporal precision, aggregates pooled tap times into consensus beat annotations, and computes intrinsic \ifdefined\GlobalTapPreprint quality \else confidence \fi measures that flag
perceptually difficult excerpts without requiring a reference.
\ifdefined\GlobalTapPreprint
In a separate study on a different 240-excerpt validation set, tapped by an
additional 733 participants,
\else
In a separate study on a different 240-excerpt validation set, tapped by an
additional 752 participants,
\fi an independent group of 130 listeners rated how
well the resulting beat tracks lined up with the music,
and we detected no overall difference between our annotations and existing
reference annotations. Crowd-derived annotations occasionally diverged from the references in phase
and metrical interpretation, yet listeners did not consistently prefer the
reference versions, suggesting that
crowd-derived and existing reference annotations can reflect different
perceptual or structural interpretations. The disagreements were
often systematic rather than random, and metrical-level choices
differed across the ten recruitment countries.
We release GlobalTap together with open-source pipeline code as a
listener-derived resource for evaluating beat-tracking systems on
culturally diverse music.
\end{abstract}

\begin{figure*}[!t]
  \centering
  \makebox[\textwidth][c]{%
    \includegraphics[alt={Pipeline schematic with five labeled stages: online tapping, pool responses, KDE estimation, consensus annotations, listener validation},width=\OverviewSchematicWidth,trim={20pt 8pt 20pt 8pt},clip]{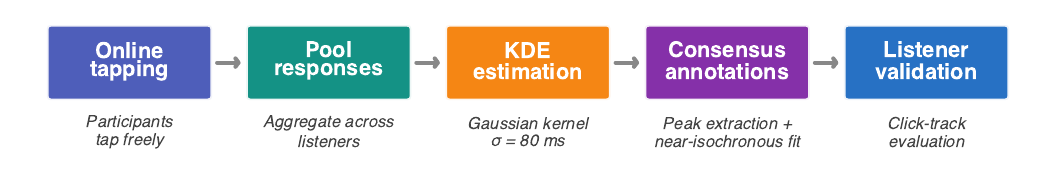}%
  }\\[-6pt]
  {\footnotesize\textbf{(A)} Crowdsourced beat-tracking pipeline.}\\[1pt]
  \makebox[\textwidth][c]{%
    \includegraphics[alt={World map showing 10 country panels of tappers (US, FR, MX, KR, EG, JP, BR, ZA, AU, PL) and 59 source countries from which the 200 benchmark excerpts were drawn},width=\OverviewMapWidth,trim={6pt 5pt 6pt 5pt},clip]{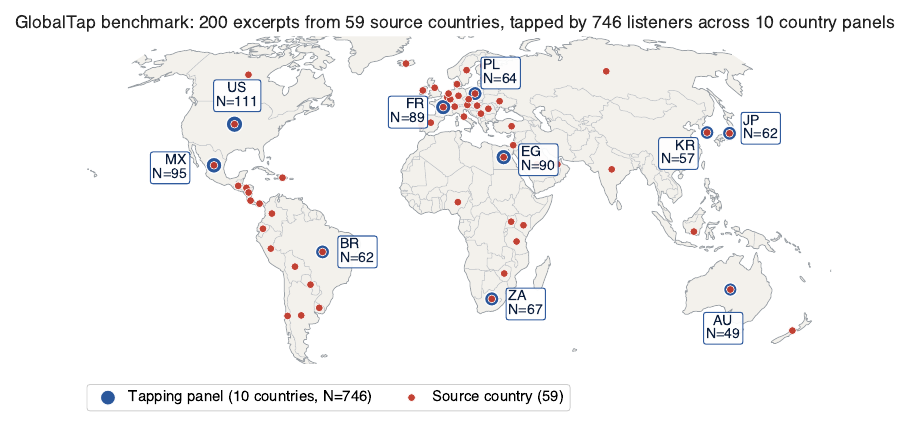}%
  }\\[-6pt]
  {\footnotesize\textbf{(B)} Geographic distribution of tapping panels and source countries.}
  \setlength{\abovecaptionskip}{2pt}
  \caption{GlobalTap pipeline overview and benchmark geography.
    \textbf{(A)} Online tapping responses are pooled across listeners,
    smoothed by Gaussian kernel-density estimation, and converted into
    a near-isochronous consensus beat annotation. A separate \ifdefined\GlobalTapPreprint listening
    experiment \else listener
    study \fi validates the pipeline on click-track overlays. \textbf{(B)} The 200 benchmark
    excerpts are drawn from 59 source countries (red). Tappers were
    recruited across 10 country panels (blue, $N=746$ tappers in total).
   The 10 countries were: United States (US), South Korea (KR), France (FR), Mexico (MX), Egypt (EG), Australia (AU), Brazil (BR), Japan (JP), Poland (PL), South Africa (ZA).
    }
  \label{fig:overview}
\end{figure*}

\section{Introduction}

Beat tracking, the automatic estimation of beat times in a musical
recording, is a core MIR task, as beat estimates support
analysis of tempo, meter, and structure \cite{davies2021tutorial}. Recent systems achieve strong
accuracy by training on multiple annotated corpora, spanning solo
instrument recordings, music with time-signature changes, and classical
music with substantial tempo variation \cite{foscarin2024beatthis}.

The community currently relies on a small set of established
annotated datasets, many of which are consolidated in the Beat This!\
unified training set \cite{foscarin2024beatthis}. This resource
comprises 4{,}556 annotated tracks from 15 corpora, including
Ballroom \cite{krebs2013rhythmic}, RWC \cite{goto2002rwc}, Beatles
\cite{harte2010beatles}, and Harmonix \cite{nieto2019harmonix}. An important limitation of these datasets, however, is that they are
heavily biased toward Western popular and dance genres
\cite{holzapfel2018ethical}. Expanding genre coverage is difficult
because traditional annotation workflows do not scale easily, since they
require recruiting musicians with expertise in the relevant musical
styles and using specialist software for time-consuming manual
annotation
\cite{rocamora2015candombe,srinivasamurthy2014supervised,foscarin2024beatthis}.

We address this limitation with a scalable online tapping pipeline
(Fig.~\ref{fig:overview}A).
Participants recruited through standard crowdsourcing platforms tap
along to musical excerpts using \href{https://gitlab.com/computational-audition/repp}{REPP},
an audio-based method for high-precision online sensorimotor-synchronization
experiments \cite{angladatort2022repp}. Pooled tap times are smoothed into a
density over perceived beat locations, and a simple optimization
step converts this density into a regularized consensus beat
annotation. The approach builds on tapping studies of beat
perception \cite{repp2005sensorimotor} and on listener-derived annotations used
in MIREX-style evaluation \cite{mckinney2007evaluation}, extended to larger and
more diverse participant pools
\cite{niarchou2022gwas,jacoby2024commonality}. Because multiple
participants respond to the same stimulus, the pooled taps also
capture variation in perceived beat location and metrical
interpretation \cite{mckinney2006ambiguity}, yielding intrinsic
\ifdefined\GlobalTapPreprint quality \else confidence \fi measures that do not require reference beats.

We validate the pipeline in a \ifdefined\GlobalTapPreprint listening experiment \else listener-preference study \fi over 240 excerpts, 200 of which are drawn from corpora with existing reference annotations (Ballroom, MIREX, and a set of mixed annotated corpora). %
We use the pipeline to create and release \textbf{GlobalTap}, a
listener-derived beat-tracking benchmark of 200 popular-music excerpts
sampled from 59 countries \cite{lee2025globalmood} and annotated by participants
recruited across ten countries (Fig.~\ref{fig:overview}B). The release includes consensus beat annotations,
de-identified tap onsets, tap-density summaries, excerpt metadata, intrinsic \ifdefined\GlobalTapPreprint quality \else confidence \fi measures, and
scripts for reproducing the aggregation and evaluation pipeline.\footnote{Code, data, and \suppl{}: \url{\RepoURL}.}

\subsection{Beat Annotation Datasets}
Existing beat-annotation corpora differ primarily in how beat times are
obtained and how much expert correction is applied. Many established
corpora use \emph{semi-automatic expert correction}, in which initial beat
estimates from an algorithm, click/metronome track, or bar-level
annotation are then checked and corrected by trained annotators.
This includes Beatles \cite{harte2010beatles}, GTZAN
\cite{tzanetakis2002musical, marchand2015gtzanrhythm}, RWC \cite{goto2002rwc,goto2006aist}, and
Ballroom \cite{krebs2013rhythmic}. TapCorrect
\cite{driedger2019tapcorrect} partly automates this step by snapping tapped
beats to nearby audio cues.

Other resources rely on manual expert annotation. Candombe
\cite{rocamora2015candombe} provides expert beat and downbeat annotations, and Harmonix \cite{nieto2019harmonix} was annotated by professional musicians in a digital audio workstation (DAW).

A related strand derives beat times from tapping or clapping
responses. Hainsworth \cite{hainsworth2004particle} used recordings
of a trained musician clapping in time with the music, while
Carnatic and Turkish makam resources used expert tapping followed by
manual correction
\cite{srinivasamurthy2014supervised,srinivasamurthy2014turkish,holzapfel2014tracking}. The
MIREX beat-evaluation set \cite{mckinney2007evaluation} goes further,
recruiting 40 listeners (approximately balanced between musicians and
non-musicians) to tap along to musical excerpts, treating each listener's taps as an individual annotation.

Finally, some datasets derive annotations from symbolic or recording
metadata rather than from audio. ASAP \cite{foscarin2020asap} aligns
scores to performances to obtain beat and downbeat times, GuitarSet \cite{xi2018guitarset} derives beats from
click-track and meter metadata, and Groove MIDI \cite{gillick2019groove}
provides tempo-aligned MIDI drum performances.

A key limitation is that most established corpora focus
on Western popular, dance, and classical genres, or on solo-instrument and
drum-set performance. Non-Western traditions appear only in a
small number of expert-built datasets, such as the Candombe
\cite{rocamora2015candombe}, Carnatic and Turkish makam
\cite{srinivasamurthy2014supervised}, and Cretan traditional music
\cite{holzapfel2014tracking} datasets used in our validation
experiment. Annotator diversity is also limited, with many corpora
relying on one or a few expert or musically trained annotators. Such
annotations can validly represent structurally informed or culturally
situated interpretations, but they do not necessarily show where ordinary listeners hear the beat.

\subsection{Tapping as a Measure of Beat Perception}
\label{sec:tapping-as-beat-perception}
Tapping along with music has long been used in music-cognition research to study beat perception, revealing how tapping behavior is shaped by tempo, meter, and musical structure \cite{repp2005sensorimotor,repp2013sensorimotor}.
Listeners do not simply tap where the onsets are, as perceived beat timing can shift depending on a sound's attack and duration, producing P-center effects
\cite{morton1976perceptual, danielsen2019whereisbeat, london2019comparison}.
Moreover, the same excerpt can support multiple metrical interpretations, so
collecting taps from many listeners provides an operational measure of
perceptual consensus, while disagreement reveals metrical ambiguity
\cite{mckinney2006ambiguity}. How easily a pulse is found depends on the music
itself, with syncopation and the amount of musical information predicting
pulse-finding ability \cite{toiviainen2003tapping}. Multiple tap sequences can
also be used to estimate both the underlying beat times and the uncertainty of
the taps around them, without ground truth \cite{dannenberg2009tap}. Tapping therefore provides a
psychologically grounded measure of beat perception, capturing when listeners
synchronize with music and the learned biases that shape how they organize
rhythmic structure \cite{sadakata2006bayesian,jacoby2017integer}.

\begin{figure}[!t]
  \centering
  \includegraphics[alt={Four stacked panels showing how audio is converted into a beat grid: an audio spectrogram, a per-participant tap raster of 40 tappers, the pooled tap-density KDE with extracted consensus peaks, and the final optimized beat grid overlaid on the audio envelope alongside the expert reference annotation},width=\columnwidth,trim={0 13mm 0 0}]{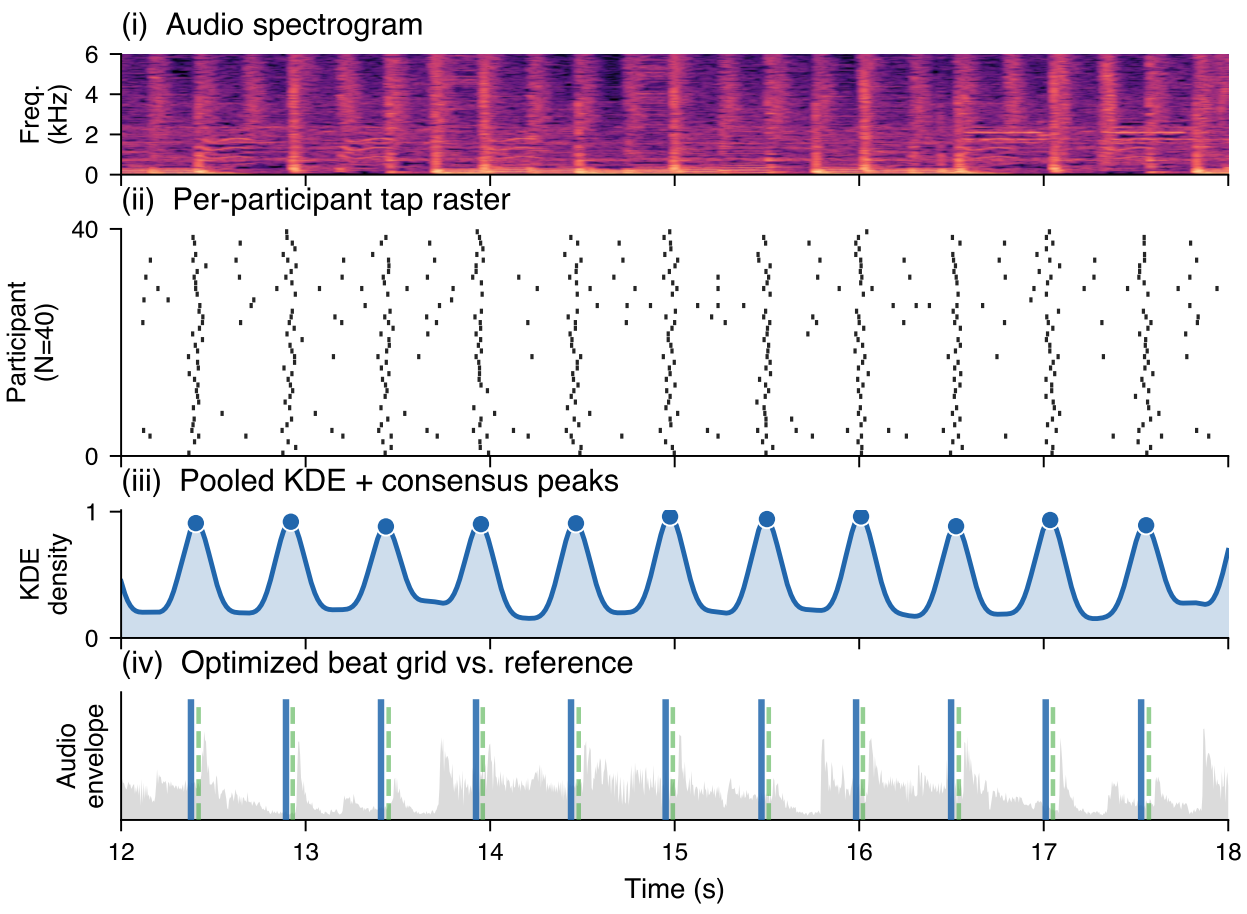}
  \setlength{\abovecaptionskip}{1pt}
  \caption{From audio to validated beat grid (Ballroom example, 12--18\,s window).
    \textbf{(i)} Audio spectrogram. \textbf{(ii)} Per-participant tap raster
    ($N{=}40$ tappers). \textbf{(iii)} Pooled tap density (Gaussian kernel
    density estimate, KDE; $\sigma{=}80$\,ms) with extracted consensus
    peaks. \textbf{(iv)}
    Optimized near-isochronous consensus grid (solid blue) overlaid on the
    audio envelope (gray) and compared with the expert reference
    annotation (dashed green).}
  \label{fig:pipeline-example}
\end{figure}

\section{Methods}

\subsection{Materials}
\label{sec:materials}

Our pipeline converts crowdsourced tapping responses into consensus
beat annotations in three stages: (a)~online tapping data collection,
(b)~kernel-density aggregation of pooled taps into an empirical
distribution over perceived beat locations, and (c)~a simple
grid-search fitting step that distills this density into a regularized
beat track (Fig.~\ref{fig:pipeline-example}). The aim is to test whether
crowdsourced tapping can support beat annotation with a transparent
proof-of-concept estimator, rather than a highly optimized audio-informed
model.

The GlobalTap benchmark contains 200 popular-music
excerpts of approximately 30\,s sampled from the 1{,}180-song GlobalMood
corpus \cite{lee2025globalmood} (see Fig.~\ref{fig:overview}B). GlobalMood
drew tracks from 2017--2023 weekly YouTube charts across 59 countries, excluding
any track charting in more than one.
The benchmark was designed to align with
our participant-recruitment strategy. Five countries
(Mexico, the United States, France, South Korea, and Egypt; see
\S\ref{sec:participants}) were oversampled for \ifdefined\GlobalTapPreprint stimuli and, on average, for listeners.\else both stimuli and listeners.\fi{} Each of these countries
contributed 20 excerpts, yielding 100 excerpts in
total. The remaining 100 excerpts were drawn at random from the other 54
countries, one or two per country, which keeps geographic coverage broad while complementing the five focal countries.

A separate 240-excerpt \emph{validation set} was used to evaluate
the pipeline against existing beat references and to test
generalization across diverse musical styles. This set comprised 80
Ballroom excerpts \cite{krebs2013rhythmic}, 20 MIREX excerpts
\cite{mckinney2007evaluation}, 100 excerpts from 13 mixed annotated corpora (\MixedCorporaDetails), and 40 additional GlobalMood excerpts separate from the benchmark. Reference annotations were available for Ballroom, MIREX,
and the mixed corpora, yielding 200 excerpts with paired Crowd and Reference conditions. The
40 GlobalMood excerpts, for which no reference annotations were
available, served as an absolute-quality check on globally
distributed popular music.

\subsection{Participants and Data Collection}
\label{sec:participants}

\textbf{Tapping participants.} \ifdefined\GlobalTapPreprint
The tapping sample comprised
$N=1{,}479$ participants, $746$ for the GlobalTap benchmark and $733$ for the
validation set, who contributed 36{,}625 trials and 2{,}185{,}514 parsed
taps.
\else
The tapping sample comprised
$N=1{,}498$ participants, $746$ for the GlobalTap benchmark and $752$ for the
validation set, who contributed 36{,}970 trials and 2{,}205{,}932 parsed
taps.
\fi For the
GlobalTap benchmark, tappers were recruited across ten country
panels (US, KR, FR, MX, EG, AU, BR, JP, PL, ZA; median 66 per
panel, range 49--111), with larger samples on average
in the five countries that were also focal source
countries for the GlobalMood stimuli (MX, US, FR, KR, EG).
\ifdefined\GlobalTapPreprint
Across panels, participants reported relatively low musical
expertise (median $1$ year of formal training; abbreviated
Gold-MSI \cite{mullensiefen2014gold} $M=4.10$ on a 1--7 scale) and broadly
comparable demographic profiles ($M_\text{age}=33.1$; $49.5\%$ female).
\else
Across panels, participants reported relatively low musical
expertise (median $1$ year of formal training; abbreviated
Gold-MSI \cite{mullensiefen2014gold} $M=4.45$ on a 1--7 scale) and broadly
comparable demographic profiles ($M_\text{age}=34.8$; $52.7\%$ female).
\fi
\ifdefined\GlobalTapPreprint
Recruitment and exclusion counts and further demographic details are reported
in \ParticipantsSuppl{}.
\else
Per-panel breakdowns, recruitment, and exclusion counts are reported in
\ParticipantsSuppl{}.
\fi

\textbf{Tapping collection.} \ifdefined\GlobalTapPreprint
All data collection was conducted
online via PsyNet \cite{harrison2020gibbs}, with participants recruited through
Prolific (and through Cint for 58 of the 296 GlobalMood validation tappers),
under an approved ethical protocol with informed consent.
\else
All data collection was conducted
online via PsyNet \cite{harrison2020gibbs} with Prolific under an approved ethical protocol with
informed consent.
\fi Tapping data were collected with \href{https://gitlab.com/computational-audition/repp}{REPP}
 (Rhythm
ExPeriment Platform), a browser-based toolkit for
sensorimotor-synchronization experiments
\cite{angladatort2022repp}, which records the stimulus and the participant's
taps in a single audio stream and recovers tap onsets by aligning that
recording to the known stimulus. This bypasses browser timer jitter and gives
precise tap timing on ordinary consumer laptops, and it has been used at scale
in prior online tapping work
\cite{niarchou2022gwas,jacoby2024commonality}. Participants
completed audio-volume calibration, a marker-recording check, and
practice trials. Those who failed these checks did not proceed to the main task.
Trials with failed recordings, failed marker analyses, or
incomplete responses were excluded. On each trial, participants
heard a ${\sim}30$\,s excerpt \ifdefined\GlobalTapPreprint (${\sim}60$\,s for
GlobalMood validation, first $30$\,s used) \fi and tapped along to the perceived
beat. Instructions read: ``Keep your taps simple, steady and evenly
spaced. Adjust if the tempo changes.'' We did not assign a
metrical level, so participants could tap the pulse they heard rather than one
we imposed \cite{mckinney2006ambiguity}.

\textbf{Validation experiment.} A separate \ifdefined\GlobalTapPreprint listening \else listener-rating
\fi experiment validated the resulting click tracks. We recruited
$N=130$ approved participants who contributed $6{,}487$
ratings across the $240$ validation excerpts (median $15$ ratings per
excerpt $\times$ condition). \ifdefined\GlobalTapPreprint
Participants rated click-track overlays on
a 7-point scale. An instruction asked them to rate
how well the clicks line up with the beat of the music, and the rating question
read ``How well did the clicks match the beat of the music?''
\else
Participants rated click-track overlays on
a 7-point scale, with the prompt ``How well do the clicks line up
with the beat of the music?''.
\fi We chose this wording because it is simple
and does not point participants toward a specific metrical level or a ``correct'' answer, and is similar to the Beat Alignment Test
\cite{iversen2008bat}. Participants judged the rendered beat sequence as a
whole and gave one overall rating, so this design cannot determine whether a
score reflects metrical level, phase, local timing, or their combination.
Two conditions were compared per
stimulus: \emph{Crowd} (beat track produced by our pipeline from
consensus tapping; see \S\ref{sec:optimization}) and
\emph{Reference} (the reference annotation for each corpus).
For Ballroom \cite{krebs2013rhythmic} and the mixed corpora these were the
corpora's own expert annotations. For MIREX, we produced a consensus sequence
by running our pipeline on the 40 MIREX lab annotators' tap sequences, so that
any Crowd--Reference difference on MIREX reflects \emph{tap source} (online
crowd vs.\ lab listeners) rather than how the taps were aggregated. Each excerpt's Crowd and Reference versions were presented in randomized
order within the same block. A multiple-choice attention check at the end of
the task asked participants what had been added on top of the music, and
identifying ``short click sounds overlaid on the piece'' was required to finish.

\subsection{Consensus Pipeline and Rendering}
\label{sec:optimization}

\textbf{Aggregation.}
Fig.~\ref{fig:pipeline-example} illustrates the aggregation,
optimization, and reference-comparison steps on one Ballroom excerpt.
We pooled individual tap times across
participants and computed an initial kernel density estimate (KDE)
on a $5$\,ms grid using a Gaussian kernel with $\sigma=80$\,ms.
Before pooling, we cleaned each participant's trial by
removing a tap when either neighboring inter-onset interval (IOI) exceeded the
trial's median IOI by more than $3.5$ scaled median absolute deviations
(MADs). The remaining taps were pooled to compute
the KDE. Peaks were extracted with prominence $\geq 0.10$\ifdefined\GlobalTapPreprint{} of the KDE maximum\fi{} and a
$150$\,ms minimum inter-peak distance \cite{virtanen2020scipy}.
The bandwidth value of $\sigma=80$\,ms was selected during pipeline development, and values between $70$ and $100$\,ms gave comparably stable period and beat-count behavior.

\ifdefined\GlobalTapPreprint
\textbf{Grid optimization.} We used an exhaustive grid-search heuristic designed for short excerpts. An independent group of 130 listeners later rated its beat grids on par with the reference annotations (\S\ref{sec:overall_validation}).
It fits one near-isochronous beat grid $\mathbf{b}=\{\phi+nT\}$ to the
extracted peaks $\mathbf{p}$ in four steps (settings in
\ApxRef{Table}{tab:app-optimizer}). First, a grid search over period $T$
($48$--$214$\,BPM) and phase $\phi$ maximized
$\mathcal{S}(T,\phi)=\bar{D}(\mathbf{b})+0.5\,C(\mathbf{b},\mathbf{p})$,
which rewards beats that land on peaks ($\bar{D}$) and grids that leave few
peaks unmatched ($C$). Here $\bar{D}$ is the mean, over the grid beats, of a
curve with the same $80$\,ms Gaussian at each peak, so every peak counts equally
however many people tapped it, and $C$ is the proportion of peaks near a grid
beat. Second, the best grid was refined by least squares on nearby peaks, and
each beat near a peak moved part of the way toward it, so the beats follow the
taps while staying close to evenly spaced.
Third, the beat rate was doubled when peaks also fell at most of the
midpoints between beats, and gaps were filled at the grid spacing. Fourth,
when the peaks formed two groups at different phases, a grid was also fit to each group, and the three
grids were compared with a second score that also rewards even beat spacing.
The grid spans the
whole excerpt, including stretches with few taps, while ratings and tracker
scores use only the $10$--$25$\,s window.
\else
\textbf{Grid optimization.} We used an exhaustive grid-search heuristic designed for short excerpts. Although the procedure could be further refined, it performed well when evaluated on an independent group of participants.
A single near-isochronous beat grid
$\mathbf{b}=\{\phi+nT\}$ was fit to the density by grid search over
period $T\in[0.28,1.25]$\,s ($48$--$214$\,BPM, $120$ steps) and
phase $\phi\in[0,T)$ ($240$ steps), scoring each candidate as
$\mathcal{S}(T,\phi)=\bar{D}(\mathbf{b})+0.5\,C(\mathbf{b},\mathbf{p})$,
where $\bar{D}$ is the mean normalized KDE value at grid beats
$\mathbf{b}$ and $C$ is the proportion of KDE peaks $\mathbf{p}$
matched by some grid beat within $0.12T+20$\,ms. The highest-scoring
grid was refined by least squares on locally matched peaks, with
each beat snapped to the nearest peak within $\leq 0.12T$. The
refined grid was post-processed by merging beats closer than
$0.6T$, inserting beats in gaps $>1.4T$, and extending the grid
forward and backward by $T$ if the local KDE supported additional
beats at either end. When the peaks split into two groups at different phases
(the largest gap between their inter-peak intervals exceeds $0.10T$), we fit a
grid to each group and kept the one with the higher score $\mathcal{S}$
\cite{mckinney2006ambiguity}.
\fi \ifdefined\GlobalTapPreprint These settings were chosen during pipeline
development on the validation corpora (Ballroom, the mixed corpora, and
MIREX), then held fixed for all reported analyses, including those of the
GlobalTap corpus. To check how much the output depends on them, we varied each
\else These
settings were selected during pipeline development using sensitivity analyses
on the validation corpora (Ballroom, the
mixed corpora, and MIREX), then held fixed for all reported analyses,
including those of the GlobalTap corpus. We varied each
\fi setting one at a time by $18$--$50$\% (e.g.\ the grid
resolution, the density/coverage weighting, the snap tolerance, or the
bimodal-detection threshold) across all 240 validation excerpts and found that outputs changed
little, \ifdefined\GlobalTapPreprint matching the original outputs \else still matching our reported beat grids \fi at F-measure $.947$--$1.000$,
with the tactus (the metrical level the grid locked onto) unchanged in
$92.1$--$100$\% of cases\ifdefined\GlobalTapPreprint, and agreement with the reference annotations changed by at most $.014$\fi.

\ifdefined\GlobalTapPreprint
\textbf{Sample size.}
\else
\textbf{Sample size selection.}
\fi \ifdefined\GlobalTapPreprint We assessed convergence with three measures: the
correlation between participant-normalized KDEs (density correlation), the F-measure between their
extracted peaks ($\pm70$\,ms, the standard beat-tracking tolerance
\cite{raffel2014mireval}), and the F-measure between the final beat grids.
\else We assessed convergence with two
measures, the correlation between participant-normalized KDEs and the
F-measure between their extracted peaks ($\pm70$\,ms, with the offset between
matched peaks as a timing check).
\fi
Across all 200 benchmark excerpts, two independent groups of $N=50$ tappers achieved a mean density correlation of $r=.814$ and a peak F-measure of $.880$, and their optimized beat grids agreed at $F=.918$. \RobustnessDetails{} evaluated smaller samples by
\ifdefined\GlobalTapPreprint
drawing two independent groups of participants per excerpt
\else
repeatedly drawing independent groups of participants
\fi and running the complete pipeline on each. \ifdefined\GlobalTapPreprint
At $N=20$, the two groups' beat grids agreed at $F=.893$.
Every validation excerpt had more tappers than that, at least $34$ (median $40$ for Ballroom, $36$ for the mixed corpora, $67$ for MIREX, and $110$ for GlobalMood validation, see \DemographicsDetails).
\else
Even groups as small as $N=20$ yielded reliable performance, with an F-measure of $.893$. Based on these results, we recruited approximately $40$ tappers per excerpt for the 240-excerpt validation set (see \DemographicsDetails).
\fi For the final benchmark we recruited 114--120 tappers per excerpt (median 117), the maximum feasible within the data-collection period.

\textbf{Rendering.} For the listening experiment, beat times in the
$10$--$25$\,s window of each excerpt were converted to $1100$-Hz,
$12$-ms Hann-windowed clicks and mixed with the original mono
\ifdefined\GlobalTapPreprint
$44.1$-kHz audio at music $0.35$ / click $0.65$, with the music
loudness-normalized to $-23$\,LUFS and the clicks at a fixed peak level.
\else
$44.1$-kHz audio, each signal peak-normalized and summed at music
$0.35$ / click $0.65$.
\fi

\subsection{Intrinsic Quality Measures}
\label{sec:intrinsic-defs}

We report two reference-free indicators of annotation quality.
Beat-sequence \emph{CV} is the coefficient of variation (SD/mean)
of the inter-onset intervals of the final beat track, a measure of
grid \emph{anisochrony} that can be computed for both Crowd and
Reference annotations. \emph{ISTC} (inter-subject tapping
coherence; \cite{singer2023coherence}) summarizes agreement among
\ifdefined\GlobalTapPreprint pre-optimization \else raw, pre-optimization \fi participant taps following the bin-counting
definition of Singer et~al. For each excerpt we count how many \emph{unique
participants} tap in each $100$\,ms bin (advanced in $50$\,ms steps), take the
largest count in each $1$\,s window, and average those maxima over the
$[0,30]$\,s window.
Because
unique-participant counts are bounded by panel size, ISTC's
absolute scale varies across corpora, so we report it
within-corpus. Both ISTC and CV are computed over the whole excerpt, while
ratings and tracker scores use the rendered $10$--$25$\,s window. ISTC is defined only for crowd-derived annotations,
because the Reference condition consists of a single annotated grid
rather than a set of taps.

\section{Results}

\subsection{KDE--Reference Alignment}
\label{sec:kde-alignment}

As a first check, we asked whether the pooled tapping density
already aligns with the reference beat structure before any grid
postprocessing is applied. Averaged over excerpts,
\ifdefined\GlobalTapPreprint
$90.3\%$ of raw KDE peaks fall within $\pm70$\,ms of a
reference beat on Ballroom, $56.2\%$ on the mixed corpora, $74.2\%$ on MIREX,
and $71.7\%$ overall.
\else
$88.3\%$ of raw KDE peaks fall within $\pm70$\,ms of a
reference beat on Ballroom, $47.5\%$ on the mixed corpora, $53.2\%$ on MIREX,
and $64.4\%$ overall, where $\pm70$\,ms is the tolerance used for standard
beat-tracking F-measure \cite{raffel2014mireval}.
\fi Agreement before grid fitting was lowest on
the mixed-corpus excerpts dominated by rubato (ASAP) or dense syncopation
(Candombe). The pooled tapping signal thus already carries much of the
temporal structure that the grid fitting then regularizes.

\subsection{Overall Listener Preference Validation}
\label{sec:overall_validation}

To test whether the optimized consensus beat track is perceived as
a viable alternative to expert annotations, we compared the
pipeline's crowd-derived track (Crowd) against each excerpt's
reference annotation. Per-excerpt mean ratings on
a 1--7 scale (see Table~\ref{tab:main}) were compared with
Wilcoxon signed-rank tests on excerpt-level means, with
$\Delta = \mathrm{Crowd} - \mathrm{Ref}$ as the effect direction.

Table~\ref{tab:main} summarizes the results. Across the 200 paired excerpts, we found no overall difference in ratings between the two annotation types
 ($\Delta=-0.01$, 95\% CI $[-0.12,+0.10]$, $p=.231$, $d_z=-0.02$), indicating no systematic listener preference for either annotation. Ratings also did not differ significantly within the Mixed or MIREX corpora\ifdefined\GlobalTapPreprint.\else, although the Mixed corpus showed a nonsignificant trend favoring the Crowd annotations, whereas MIREX showed a nonsignificant trend favoring the Reference annotations.\fi{} Reference annotations received significantly higher ratings in the Ballroom corpus, but the effect size was small. These results suggest that our Crowd-based annotations, as judged by the independent raters, are comparable in quality to the references.

The 40 reference-free GlobalMood excerpts received plausible ratings ($M=4.93$), comparable to those observed for the other corpora. Mean ratings across corpora ranged from $4.33$ to $5.14$\ifdefined\GlobalTapPreprint{} on the 7-point scale.\else, below the maximum possible rating of $7$. These moderate absolute ratings are expected because participants evaluated each excerpt relative to the other stimuli presented in the experiment, which tends to pull scores toward the middle of the scale. Substantial individual differences in scale use may have further lowered the means.\fi

\begin{table}[!t]
\centering
\caption{Per-corpus mean listener ratings (1--7 scale) for the
  pipeline (Crowd) vs.\ Reference annotations on the 200 paired
  validation excerpts plus the 40-excerpt GlobalMood absolute-quality
  set. $\Delta=\mathrm{Crowd}-\mathrm{Ref}$; effect sizes are
  paired-sample Cohen's $d_z$; $p$-values from Wilcoxon signed-rank
  on excerpt-level means. %
  }
\label{tab:main}
\small
\setlength{\tabcolsep}{4pt}
\noindent\resizebox{0.96\columnwidth}{!}{%
\begin{tabular}{lrrrrrr}
\toprule
Corpus            & $n$ & Crowd & Ref  & $\Delta$ & $p$  & $d_z$ \\
\midrule
Overall (paired)  & 200 & 4.67 & 4.68 & $-0.01$ & .231 & $-0.02$ \\
Ballroom          &  80 & 5.01 & 5.14 & $-0.13$ & .007 & $-0.22$ \\
Mixed corpora     & 100 & 4.47 & 4.37 & $+0.09$ & .546 & $+0.10$ \\
MIREX             &  20 & 4.33 & 4.41 & $-0.08$ & .557 & $-0.09$ \\
GlobalMood        &  40 & 4.93 & ---  & ---     & ---  & ---    \\
\bottomrule
\end{tabular}%
}
\end{table}

\subsection{Intrinsic Quality Measures Correlate with Listener Ratings}
\label{sec:intrinsic}

Both reference-free measures were associated with ratings, supporting their
use as reliability indicators. \ifdefined\GlobalTapPreprint The Crowd
annotation's beat-sequence CV correlated with both Crowd \else Beat-sequence CV
correlated with Crowd \fi(Spearman $\rho=-.43$, $p<.0001$) and Reference ratings
\ifdefined\GlobalTapPreprint
($\rho=-.39$, $p<.0001$) across the 200 paired excerpts, so temporally
irregular excerpts seem harder to annotate, whatever the method.
\else
($\rho=-.39$, $p<.0001$) across the 200 paired excerpts, consistent with
temporally irregular stimuli being genuinely harder, not merely harder for the
pipeline.
\fi Within individual
corpora, CV also predicted Crowd ratings for the mixed corpora ($\rho=-.53$,
$p<.001$) and Ballroom ($\rho=-.35$, $p=.0017$), \ifdefined\GlobalTapPreprint but not for GlobalMood ($\rho=-.23$, $p=.15$) or MIREX ($\rho=-.02$, $p=.94$).\else while MIREX did not reach
significance ($\rho=-.02$, $p=.94$).\fi{} Pre-optimization
ISTC correlated positively with Crowd ratings in the three larger corpora:
\ifdefined\GlobalTapPreprint
mixed ($\rho=.68$, $p<.0001$), Ballroom ($\rho=.40$, $p=.0002$), and
GlobalMood ($\rho=.52$, $p=.0006$), but not for the smaller MIREX set.
All significant associations remain significant after Benjamini--Hochberg
correction.
\else
mixed ($\rho=.67$, $p<.0001$), Ballroom ($\rho=.40$, $p=.0003$), and
GlobalMood ($\rho=.45$, $p=.004$), but not for the smaller MIREX set.
\fi

\subsection{GlobalTap Benchmark Analyses}
\label{sec:benchmark_analyses}

On GlobalTap, pooled crowd consensus from \ifdefined\GlobalTapPreprint all 746 benchmark tappers (\S\ref{sec:participants}) agreed with madmom
\cite{bock2014multimodel}\else the complete 746-participant
recruitment panel (\S\ref{sec:participants}) agreed with madmom
\cite{bock2016joint}\fi{} at $F=.897$ and
Beat This!\ \cite{foscarin2024beatthis} at $F=.841$ ($\pm70$\,ms), and
tracker--tracker agreement was $F=.875$. These values and the metrics below
use the 198 excerpts evaluable for both trackers (Beat This!\ returned fewer
than two beats for two excerpts).
Crowd--tracker agreement was thus comparable to
tracker--tracker agreement, supporting GlobalTap's direct use as a
listener-derived evaluation reference.
Using the GlobalTap consensus as the reference, we also computed the
standard beat-tracking evaluation metrics
\cite{davies2009evaluation}, including correct metric level (CML) and any
metric level (AML, which credits double, half, and offbeat interpretations),
each over the longest continuously correct stretch (CMLc/AMLc) or all beats
(CMLt/AMLt), along with information gain, which measures how far the spread of
beat errors departs from uniform.
All were computed with mir\_eval defaults \cite{raffel2014mireval}.
Detailed results are reported in \BeatMetricsDetails. Both methods achieved high overall scores, although madmom systematically outperformed Beat This! on all five metrics. \ifdefined\GlobalTapPreprint Most of madmom's advantage
disappears when other metrical levels and offbeat interpretations are accepted
(CMLt $.807$ vs.\ $.636$; AMLt $.956$ vs.\ $.935$), so it lies mainly in which
level and phase each tracker chose.
\else Much of madmom's advantage on this benchmark appears to reflect more accurate selection of the metrical level, rather than more precise beat timing within a given level.
\fi %

\subsection{Exploratory Country-Panel Analyses}
\label{sec:country_analyses}

\begin{figure}[!t]
  \centering
  \includegraphics[alt={Four panels, one per period relation (same rate, double rate, half rate, other), each showing the percentage of trials in that relation for the ten country panels with 95 percent cluster-bootstrap intervals and a dashed line for the mean across panels},width=\CountryPanelWidth]{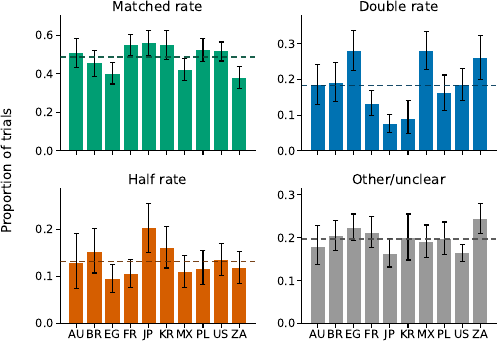}
  \vspace{-14pt}
\ifdefined\GlobalTapPreprint
  \caption{Metrical-level choice by country panel. Proportion of trials whose
    tapping rate matched, doubled, halved, or otherwise diverged from the
    median of the other nine panels (95\% cluster-bootstrap intervals, dashed
    line is the mean across panels).}
\else
  \caption{Metrical-level choice by country panel. Proportion of trials whose
    tapping period matched, doubled, halved, or otherwise diverged from the
    median of the other nine panels (95\% cluster-bootstrap intervals, dashed
    line is the mean across panels).}
\fi
  \label{fig:country-panels}
\end{figure}

For each excerpt, we compared each trial's tapping period with the median
period across the other nine panels (Fig.~\ref{fig:country-panels}). Metrical-level selection differed
significantly across the ten recruitment panels overall ($p<.001$). For
example, Japan showed the least double-rate and most half-rate tapping of
any panel ($7.5\%/20.1\%$), while \ifdefined\GlobalTapPreprint Egypt, Mexico,
and South Africa tapped at double rate most often ($28.1\%$, $28.0\%$, and
$26.0\%$), \else Egypt showed the opposite extreme
($28.1\%/9.4\%$), \fi \ifdefined\GlobalTapPreprint a difference \else an interesting discrepancy \fi to explore in
future work. Tap timing (phase) showed no consistent differences across
panels ($p=.152$). In the five \ifdefined\GlobalTapPreprint focal countries, \else countries from which we sampled 20 excerpts
each, \fi tapping coherence was not reliably higher on their own country's excerpts
($+0.07$ ISTC units, 95\% CI $[-0.03,+0.16]$, $p=.18$). Note that we grouped participants by where they were recruited, which may not reflect the music they are enculturated to, and we did not measure how familiar they were with individual excerpts.

\subsection{Structured Divergences Between Crowd and Reference Annotations}
\label{sec:divergences}

When the pipeline diverged from expert annotations, the
differences were often systematic and musically interpretable. On
the 180 excerpts with paired expert beat references
(80~Ballroom and 100~mixed-corpora, with the 20 MIREX excerpts
excluded because their reference is itself a crowd-pipeline output),
three recurring patterns emerged. An example of each is in the \suppldemos{}.

\textit{Phase offsets.}
\ifdefined\GlobalTapPreprint
Consistent offsets $\geq20$\,ms appeared on $71$ of the $145$ excerpts where
the crowd and reference beat periods matched. On $8$ of them the crowd's beats fell between the
reference beats, a third to half a beat away. On the other $63$, the crowd
preceded the reference in $49$ cases (up to $144$\,ms).
\else
Consistent offsets $\geq20$\,ms appeared on $63/145$
tactus-matched excerpts. The crowd preceded the reference in $49/63$ cases
(up to $144$\,ms).
\fi This is consistent with negative mean asynchrony,
P-center effects, or attending to different instrumental or rhythmic layers,
though our design cannot separate them
\cite{repp2005sensorimotor,morton1976perceptual,danielsen2019whereisbeat,
rocamora2015candombe}.

\textit{Metrical alternatives.} On $35/180$ excerpts, the
\ifdefined\GlobalTapPreprint
crowd annotation's period differed from the reference's: 9 at half rate, 14 at
double rate, and 12 at other ratios (six near 3:1, 1:3, or 2:3).
\else
dominant tapping
period differed from the reference: 9 at half rate, 14 at double rate, and 12
at other (non-integer) ratios.
\fi The nine half-rate cases all had fast reference
beats (median IOIs $.28$--$.37$\,s) while the crowd tapped near preferred
rates ($.57$--$.75$\,s), consistent with listeners favoring a slower level at
fast tempi \cite{mckinney2006ambiguity}. \ifdefined\GlobalTapPreprint All nine
also received higher Crowd than Reference mean ratings (median $\Delta=+.75$;
$+.60$ for the five Ballroom excerpts); \else They also received higher Crowd than
Reference mean ratings (median $\Delta=+.75$), and five were Ballroom
excerpts (median $\Delta=+.60$); \fi double-rate and other cases showed no
consistent direction
(\MetricalDetails).
On the $145$ \ifdefined\GlobalTapPreprint excerpts with matching periods, \else tactus-matched excerpts \fi the reference was rated
slightly higher (median $\Delta=-.14$, Crowd higher on $51/145$). \ifdefined\GlobalTapPreprint
\else These
comparisons likely describe listener preference rather than errors in the reference
annotations.
\fi

\textit{Bimodal pooled tapping.} On $31/180$ excerpts, the KDE
\ifdefined\GlobalTapPreprint
also showed a pulse at half or double the crowd annotation's rate, with at least
$40\%$ as many inter-peak intervals at that rate as at the annotation's,
indicating that the pooled taps contained two metrical levels before the optimizer selected one grid.
\else
showed a second group of inter-peak intervals
at $\geq40\%$ the strength of the dominant one, indicating multiple
periodicities before the optimizer selected one grid.
\fi

\section{Discussion}

This study introduces GlobalTap, a fully reproducible benchmark of 200 excerpts from 59 countries, and shows that online crowd tapping provides a scalable approach to beat annotation. Validation analyses support its perceptual validity: overall, listeners rated crowd-derived annotations as comparable to existing references, and beat-sequence CV and ISTC predicted these ratings. Crowd tapping therefore offers a practical complement to expert annotation, which remains valuable but is costly, requires specialist knowledge, and is difficult to scale across diverse musical contexts.
Systematic differences between crowd-derived and reference annotations were more consistent with perceptual ambiguity or alternative interpretations than with annotation failure. \ifdefined\GlobalTapPreprint How listeners perceive and evaluate rhythm also depends partly on familiarity with culturally specific musical styles \cite{cameron2015cross,hannon2012familiarity,jakubowski2022aesthetics,jacoby2024commonality}, which motivates two features of our approach: it allows us to recruit participants in countries the music was sampled from, and it focuses on non-expert listeners, whose interpretations may differ meaningfully from those of experts, for example in the choice of tactus level \cite{drake2000tapping}, and potentially better reflect ordinary listening.\else This aligns with evidence that beat finding depends partly on familiarity with culturally specific musical styles \cite{cameron2015cross,hannon2012familiarity,jakubowski2022aesthetics,jacoby2024commonality}. Our approach thus offers two related advantages: it allows us to recruit participants in the countries the music was sampled from, and it focuses on non-expert listeners, whose interpretations may differ meaningfully from those of experts, for example in the choice of tactus level \cite{london2019tapping,mckinney2006ambiguity}, and potentially better reflect ordinary listening.\fi{} Neither source replaces the other, and non-expert listeners may still miss structure that an enculturated musician would identify.

\textbf{Limitations and future directions.}
The dataset introduced here remains limited in its geographic coverage and in the scale and density of its annotations. With only 200 excerpts, it is currently better suited for benchmarking than training deep-learning models. However, the pipeline is scalable, and the dataset could be expanded \ifdefined\GlobalTapPreprint by recruiting further through platforms such as Prolific and Cint \else by recruiting through platforms beyond Prolific, such as Cint, \fi to reach a wider set of countries \cite{van2026one}.

A further limitation is that our current beat-extraction procedure relies on simple heuristics. For example, it does not accommodate music with substantial tempo fluctuations, such as rubato, or more complex forms of rhythmic organization. This limitation could be addressed in future work, because tapping data can reflect tempo changes provided that those changes are sufficiently predictable \cite{repp1999control}. The extraction procedure could also be extended by incorporating acoustic information. Although we did not use acoustic features here, annotations derived from tapping can be refined by aligning them to audio cues \cite{driedger2019tapcorrect}, and audio-based trackers \ifdefined\GlobalTapPreprint\cite{bock2014multimodel, foscarin2024beatthis}\else\cite{bock2016joint, foscarin2024beatthis}\fi{} could supply those cues, sharpening timing while leaving the metrical level and phase to listeners. 

Despite these limitations, the approach offers a promising means of addressing important gaps in the literature. It could be used to investigate the psychological and cultural factors that shape listeners' choices of tactus and meter, while intrinsic quality measures could provide complementary insight into the difficulty or complexity of metrical structure. \ifdefined\GlobalTapPreprint Because recruitment platforms can screen for musical training, the same pipeline can also collect annotations from musicians, non-musicians, or mixed groups and compare them directly. \fi The paradigm could also be extended to produce high-quality annotations for challenging musical materials that remain beyond the capabilities of current state-of-the-art methods, for example by selectively recruiting larger numbers of tappers for tracks \ifdefined\GlobalTapPreprint where tappers agree poorly with one another or with current trackers.\else that show low agreement.\fi{} The same framework could also support the collection of other timestamped MIR annotations, including downbeats and section boundaries. Taken together, we view this work as a step toward the broader inclusion of popular musical materials from around the world in beat-tracking research and MIR more generally.

\section*{Ethics Statement}
We conducted this study in accordance with ethical best practices.
\ifdefined\GlobalTapPreprint
All participants, recruited via Prolific or Cint, provided informed consent under
Cornell University IRB protocol IRB0150361 (``The cultural foundation of
perception and cognition -- online experiments''), were compensated fairly according to the platforms'
guidelines, and had their data protected by standard safeguards.
\else
All participants, recruited via Prolific, provided informed consent under
Cornell University IRB protocol IRB0150361 (``The cultural foundation of
perception and cognition -- online experiments''), were compensated fairly according to Prolific's
guidelines, and had their data protected by standard safeguards.
\fi No personally identifying information is released. \ifdefined\GlobalTapPreprint The participant data released are limited to de-identified tap onsets, tap-density
summaries, derived consensus annotations, per-excerpt mean ratings, and intrinsic-quality measures.\else The public data release is limited to de-identified tap onsets, tap-density
summaries, derived consensus annotations, and intrinsic-quality measures.\fi{}
Full GlobalTap benchmark source recordings are not redistributed and must be
obtained from their source corpora and YouTube, as documented in the \href{\RepoURL}{repository} (\href{\SourcesURL}{\texttt{data/SOURCES.md}}). Cross-cultural research requires careful
attention to a wide range of ethical considerations
\cite{savage2023building,jacoby2020cross}, and these concerns informed the
design of the present study. We acknowledge that training and evaluation
datasets may contain biases introduced through participant recruitment,
sampling, and data-selection procedures. Nevertheless, our study aims to
address important limitations and societal risks in current MIR technologies,
which remain disproportionately focused on Western music and Western
participants \cite{holzapfel2018ethical}. \ifdefined\GlobalTapPreprint
Despite these efforts, our
benchmark panels were recruited through a single crowdsourcing platform,
Prolific, which draws mostly on Western participants, and the panels remain
uneven in size.
\else
Despite these efforts, our
listener panels were recruited through a single crowdsourcing platform,
Prolific, which draws mostly on Western participants, and the panels remain
uneven in size.
\fi The benchmark should therefore be understood as moving closer to
listener-derived evaluation across cultures.

\section*{AI Usage Statement}
During original pipeline development, AI tools were used to help run the
sensitivity analyses that informed the parameters reported in Methods. The
authors selected and verified the final values. During final revisions, these
tools also helped refine the design of additional analyses, check their
implementation, and draft documentation for the released code and
supplementary material. The authors verified all code and outputs and take
responsibility for the paper. AI tools were also used to assist manuscript
editing and proofreading. The authors reviewed each of the edit suggestions and
approved the final version.

\ifdefined\GlobalTapPreprint
  \bibliography{reference_preprint}
\else
  \bibliography{reference}

\begin{thebibliography}{10}
\providecommand{\url}[1]{#1}
\csname url@samestyle\endcsname
\providecommand{\newblock}{\relax}
\providecommand{\bibinfo}[2]{#2}
\providecommand{\BIBentrySTDinterwordspacing}{\spaceskip=0pt\relax}
\providecommand{\BIBentryALTinterwordstretchfactor}{4}
\providecommand{\BIBentryALTinterwordspacing}{\spaceskip=\fontdimen2\font plus
\BIBentryALTinterwordstretchfactor\fontdimen3\font minus \fontdimen4\font\relax}
\providecommand{\BIBforeignlanguage}[2]{{%
\expandafter\ifx\csname l@#1\endcsname\relax
\typeout{** WARNING: IEEEtran.bst: No hyphenation pattern has been}%
\typeout{** loaded for the language `#1'. Using the pattern for}%
\typeout{** the default language instead.}%
\else
\language=\csname l@#1\endcsname
\fi
#2}}
\providecommand{\BIBdecl}{\relax}
\BIBdecl

\bibitem{davies2021tutorial}
\BIBentryALTinterwordspacing
M.~E.~P. Davies, S.~B{\"o}ck, and M.~Fuentes, ``Tempo, beat and downbeat estimation,'' Tutorial at the 22nd International Society for Music Information Retrieval Conference, 2021. [Online]. Available: \url{https://tempobeatdownbeat.github.io/tutorial/intro.html}
\BIBentrySTDinterwordspacing

\bibitem{foscarin2024beatthis}
F.~Foscarin, J.~Schl{\"u}ter, and G.~Widmer, ``{Beat This!} {A}ccurate beat tracking without {DBN} postprocessing,'' in \emph{Proc. ISMIR}, 2024, pp. 962--969.

\bibitem{krebs2013rhythmic}
F.~Krebs, S.~B{\"o}ck, and G.~Widmer, ``Rhythmic pattern modeling for beat and downbeat tracking in musical audio,'' in \emph{Proc. ISMIR}, 2013, pp. 227--232.

\bibitem{goto2002rwc}
M.~Goto, H.~Hashiguchi, T.~Nishimura, and R.~Oka, ``{RWC} music database: Popular, classical, and jazz music databases,'' in \emph{Proc. ISMIR}, 2002, pp. 287--288.

\bibitem{harte2010beatles}
\BIBentryALTinterwordspacing
C.~Harte, ``Towards automatic extraction of harmony information from music signals,'' Ph.D. dissertation, Queen Mary University of London, Aug. 2010, source of the Isophonics Beatles annotations. [Online]. Available: \url{https://theses.eurasip.org/document/towards-automatic-extraction-of-harmony-information-from-music-signals/}
\BIBentrySTDinterwordspacing

\bibitem{nieto2019harmonix}
O.~Nieto, M.~McCallum, M.~E.~P. Davies, A.~Robertson, A.~Stark, and E.~Egozy, ``The {Harmonix Set}: Beats, downbeats, and functional segment annotations of {W}estern popular music,'' in \emph{Proc. ISMIR}, 2019, pp. 565--572.

\bibitem{holzapfel2018ethical}
A.~Holzapfel, B.~L. Sturm, and M.~Coeckelbergh, ``Ethical dimensions of music information retrieval technology,'' \emph{Transactions of the International Society for Music Information Retrieval}, vol.~1, no.~1, pp. 44--55, 2018.

\bibitem{rocamora2015candombe}
L.~O. Nunes, M.~Rocamora, L.~Jure, and L.~W.~P. Biscainho, ``Beat and downbeat tracking based on rhythmic patterns applied to the {Uruguayan Candombe} drumming,'' in \emph{Proc. ISMIR}, 2015, pp. 264--270.

\bibitem{srinivasamurthy2014supervised}
A.~Srinivasamurthy and X.~Serra, ``A supervised approach to hierarchical metrical cycle tracking from audio music recordings,'' in \emph{Proc. ICASSP}, 2014, pp. 5217--5221.

\bibitem{angladatort2022repp}
M.~Anglada-Tort, P.~M. Harrison, and N.~Jacoby, ``{REPP}: A robust cross-platform solution for online sensorimotor synchronization experiments,'' \emph{Behavior Research Methods}, vol.~54, no.~5, pp. 2271--2285, 2022.

\bibitem{repp2005sensorimotor}
B.~H. Repp, ``Sensorimotor synchronization: A review of the tapping literature,'' \emph{Psychonomic Bulletin \& Review}, vol.~12, no.~6, pp. 969--992, 2005.

\bibitem{mckinney2007evaluation}
M.~F. McKinney, D.~Moelants, M.~E.~P. Davies, and A.~Klapuri, ``Evaluation of audio beat tracking and music tempo extraction algorithms,'' \emph{Journal of New Music Research}, vol.~36, no.~1, pp. 1--16, 2007.

\bibitem{niarchou2022gwas}
M.~Niarchou, D.~E. Gustavson, J.~F. Sathirapongsasuti, M.~Anglada-Tort, E.~Eising, E.~Bell \emph{et~al.}, ``Genome-wide association study of musical beat synchronization demonstrates high polygenicity,'' \emph{Nature Human Behaviour}, vol.~6, no.~9, pp. 1292--1309, 2022.

\bibitem{jacoby2024commonality}
N.~Jacoby, R.~Polak, J.~A. Grahn, D.~J. Cameron, K.~M. Lee, R.~Godoy, E.~A. Undurraga, T.~Huanca, T.~Thalwitzer, N.~Doumbia, D.~Goldberg, E.~H. Margulis, P.~C.~M. Wong, L.~Jure, M.~Rocamora, S.~Fujii, P.~E. Savage, J.~Ajimi, R.~Konno, S.~Oishi, K.~Jakubowski, A.~Holzapfel, E.~Mungan, E.~Kaya, P.~Rao, M.~A. Rohit, S.~Alladi, B.~Tarr, M.~Anglada-Tort, P.~M.~C. Harrison, M.~J. McPherson, S.~Dolan, A.~Durango, and J.~H. McDermott, ``Commonality and variation in mental representations of music revealed by a cross-cultural comparison of rhythm priors in 15 countries,'' \emph{Nature Human Behaviour}, vol.~8, no.~5, pp. 846--877, 2024.

\bibitem{mckinney2006ambiguity}
M.~F. McKinney and D.~Moelants, ``Ambiguity in tempo perception: What draws listeners to different metrical levels?'' \emph{Music Perception}, vol.~24, no.~2, pp. 155--166, 2006.

\bibitem{lee2025globalmood}
H.~Lee, E.~Celen, P.~M.~C. Harrison, M.~Anglada-Tort, P.~van Rijn, M.~Park, M.~Sch{\"o}nwiesner, and N.~Jacoby, ``{GlobalMood}: A cross-cultural benchmark for music emotion recognition,'' in \emph{Proc. ISMIR}, 2025, pp. 11--19.

\bibitem{tzanetakis2002musical}
G.~Tzanetakis and P.~Cook, ``Musical genre classification of audio signals,'' \emph{IEEE Transactions on Speech and Audio Processing}, vol.~10, no.~5, pp. 293--302, 2002.

\bibitem{marchand2015gtzanrhythm}
U.~Marchand, Q.~Fresnel, and G.~Peeters, ``{GTZAN-Rhythm}: Extending the {GTZAN} test-set with beat, downbeat and swing annotations,'' in \emph{Late-Breaking/Demo Session, Proc. ISMIR}, 2015, {HAL} preprint hal-01252607.

\bibitem{goto2006aist}
M.~Goto, ``{AIST} annotation for the {RWC} music database,'' in \emph{Proc. ISMIR}, 2006, pp. 359--360.

\bibitem{driedger2019tapcorrect}
J.~Driedger, H.~Schreiber, W.~B. de~Haas, and M.~M{\"u}ller, ``Towards automatically correcting tapped beat annotations for music recordings,'' in \emph{Proc. ISMIR}, 2019, pp. 200--207.

\bibitem{hainsworth2004particle}
S.~Hainsworth and M.~Macleod, ``Particle filtering applied to musical tempo tracking,'' \emph{EURASIP Journal on Applied Signal Processing}, vol. 2004, no.~15, pp. 2385--2395, 2004.

\bibitem{srinivasamurthy2014turkish}
A.~Srinivasamurthy, A.~Holzapfel, and X.~Serra, ``In search of automatic rhythm analysis methods for {Turkish} and {Indian} art music,'' \emph{Journal of New Music Research}, vol.~43, no.~1, pp. 94--114, 2014.

\bibitem{holzapfel2014tracking}
A.~Holzapfel, F.~Krebs, and A.~Srinivasamurthy, ``Tracking the ``odd'': Meter inference in a culturally diverse music corpus,'' in \emph{Proc. ISMIR}, 2014, pp. 425--430.

\bibitem{foscarin2020asap}
F.~Foscarin, A.~McLeod, P.~Rigaux, F.~Jacquemard, and M.~Sakai, ``{ASAP}: A dataset of aligned scores and performances for piano transcription,'' in \emph{Proc. ISMIR}, 2020, pp. 534--541.

\bibitem{xi2018guitarset}
Q.~Xi, R.~M. Bittner, J.~Pauwels, X.~Ye, and J.~P. Bello, ``{GuitarSet}: A dataset for guitar transcription,'' in \emph{Proc. ISMIR}, 2018, pp. 453--460.

\bibitem{gillick2019groove}
J.~Gillick, A.~Roberts, J.~Engel, D.~Eck, and D.~Bamman, ``Learning to groove with inverse sequence transformations,'' in \emph{Proceedings of the 36th International Conference on Machine Learning}, ser. Proceedings of Machine Learning Research, vol.~97, 2019, pp. 2269--2279.

\bibitem{repp2013sensorimotor}
B.~H. Repp and Y.-H. Su, ``Sensorimotor synchronization: A review of recent research (2006--2012),'' \emph{Psychonomic Bulletin \& Review}, vol.~20, no.~3, pp. 403--452, 2013.

\bibitem{morton1976perceptual}
J.~Morton, S.~Marcus, and C.~Frankish, ``Perceptual centers ({P}-centers),'' \emph{Psychological Review}, vol.~83, no.~5, pp. 405--408, 1976.

\bibitem{danielsen2019whereisbeat}
A.~Danielsen, K.~Nymoen, E.~Anderson, G.~S. C{\^a}mara, M.~T. Langer{\o}d, M.~R. Thompson, and J.~London, ``Where is the beat in that note? {E}ffects of attack, duration, and frequency on the perceived timing of musical and quasi-musical sounds,'' \emph{Journal of Experimental Psychology: Human Perception and Performance}, vol.~45, no.~3, pp. 402--418, 2019.

\bibitem{london2019comparison}
J.~London, K.~Nymoen, M.~T. Langer{\o}d, M.~R. Thompson, D.~L. Code, and A.~Danielsen, ``A comparison of methods for investigating the perceptual center of musical sounds,'' \emph{Attention, Perception, \& Psychophysics}, vol.~81, no.~6, pp. 2088--2101, 2019.

\bibitem{toiviainen2003tapping}
P.~Toiviainen and J.~S. Snyder, ``Tapping to {Bach}: Resonance-based modeling of pulse,'' \emph{Music Perception}, vol.~21, no.~1, pp. 43--80, 2003.

\bibitem{dannenberg2009tap}
R.~B. Dannenberg and L.~A. Wasserman, ``Estimating the error distribution of a single tap sequence without ground truth,'' in \emph{Proc. ISMIR}, 2009, pp. 297--302.

\bibitem{sadakata2006bayesian}
M.~Sadakata, P.~Desain, and H.~Honing, ``The {B}ayesian way to relate rhythm perception and production,'' \emph{Music Perception}, vol.~23, no.~3, pp. 269--288, 2006.

\bibitem{jacoby2017integer}
N.~Jacoby and J.~H. McDermott, ``Integer ratio priors on musical rhythm revealed cross-culturally by iterated reproduction,'' \emph{Current Biology}, vol.~27, no.~3, pp. 359--370, 2017.

\bibitem{mullensiefen2014gold}
D.~M{\"u}llensiefen, B.~Gingras, J.~Musil, and L.~Stewart, ``The musicality of non-musicians: An index for assessing musical sophistication in the general population,'' \emph{PLOS ONE}, vol.~9, no.~2, p. e89642, 2014.

\bibitem{harrison2020gibbs}
P.~Harrison, R.~Marjieh, F.~Adolfi, P.~van Rijn, M.~Anglada-Tort, O.~Tchernichovski, P.~Larrouy-Maestri, and N.~Jacoby, ``Gibbs sampling with people,'' \emph{Advances in Neural Information Processing Systems}, vol.~33, pp. 10\,659--10\,671, 2020.

\bibitem{iversen2008bat}
J.~R. Iversen and A.~D. Patel, ``The beat alignment test ({BAT}): Surveying beat processing abilities in the general population,'' in \emph{Proceedings of the 10th International Conference on Music Perception and Cognition}, Sapporo, Japan, 2008, pp. 465--468.

\bibitem{virtanen2020scipy}
P.~Virtanen, R.~Gommers, T.~E. Oliphant, M.~Haberland, T.~Reddy, D.~Cournapeau, E.~Burovski, P.~Peterson, W.~Weckesser, J.~Bright \emph{et~al.}, ``{SciPy} 1.0: fundamental algorithms for scientific computing in {P}ython,'' \emph{Nature Methods}, vol.~17, no.~3, pp. 261--272, 2020.

\bibitem{raffel2014mireval}
C.~Raffel, B.~McFee, E.~J. Humphrey, J.~Salamon, O.~Nieto, D.~Liang, and D.~P.~W. Ellis, ``mir\_eval: A transparent implementation of common {MIR} metrics,'' in \emph{Proc. ISMIR}, 2014, pp. 367--372.

\bibitem{singer2023coherence}
N.~Singer, N.~Jacoby, T.~Hendler, and R.~Granot, ``Feeling the beat: Temporal predictability is associated with ongoing changes in music-induced pleasantness,'' \emph{Journal of Cognition}, vol.~6, no.~1, p.~34, 2023.

\bibitem{bock2014multimodel}
S.~B{\"o}ck, F.~Krebs, and G.~Widmer, ``A multi-model approach to beat tracking considering heterogeneous music styles,'' in \emph{Proc. ISMIR}, 2014, pp. 603--608.

\bibitem{davies2009evaluation}
M.~E.~P. Davies, N.~Degara, and M.~D. Plumbley, ``Evaluation methods for musical audio beat tracking algorithms,'' Queen Mary University of London, Centre for Digital Music, Tech. Rep. C4DM-TR-09-06, 2009.

\bibitem{cameron2015cross}
D.~J. Cameron, J.~Bentley, and J.~A. Grahn, ``Cross-cultural influences on rhythm processing: reproduction, discrimination, and beat tapping,'' \emph{Frontiers in Psychology}, vol.~6, p. 366, 2015.

\bibitem{hannon2012familiarity}
E.~E. Hannon, G.~Soley, and S.~Ullal, ``Familiarity overrides complexity in rhythm perception: a cross-cultural comparison of {A}merican and {T}urkish listeners,'' \emph{Journal of Experimental Psychology: Human Perception and Performance}, vol.~38, no.~3, pp. 543--548, 2012.

\bibitem{jakubowski2022aesthetics}
K.~Jakubowski, R.~Polak, M.~Rocamora, L.~Jure, and N.~Jacoby, ``Aesthetics of musical timing: Culture and expertise affect preferences for isochrony but not synchrony,'' \emph{Cognition}, vol. 227, p. 105205, 2022.

\bibitem{drake2000tapping}
C.~Drake, A.~Penel, and E.~Bigand, ``Tapping in time with mechanically and expressively performed music,'' \emph{Music Perception}, vol.~18, no.~1, pp. 1--23, 2000.

\bibitem{van2026one}
P.~van Rijn, Y.~Sun, H.~Lee, R.~Marjieh, I.~Sucholutsky, F.~Lanzarini, E.~Andr{\'e}, and N.~Jacoby, ``One test, many tongues: Surveying language proficiency across the globe,'' \emph{Proceedings of the National Academy of Sciences}, vol. 123, no.~13, p. e2420179123, 2026.

\bibitem{repp1999control}
B.~H. Repp, ``Control of expressive and metronomic timing in pianists,'' \emph{Journal of Motor Behavior}, vol.~31, no.~2, pp. 145--164, 1999.

\bibitem{savage2023building}
P.~E. Savage, N.~Jacoby, E.~H. Margulis, H.~Daikoku, M.~Anglada-Tort, S.~E.-S. Castelo-Branco, F.~E. Nweke, S.~Fujii, S.~Hegde, {Hu Chuan-Peng}, J.~Jabbour, C.~Lew-Williams, D.~Mangalagiu, R.~McNamara, D.~M{\"u}llensiefen, P.~Opondo, A.~D. Patel, and H.~Schippers, ``Building sustainable global collaborative networks: Recommendations from music studies and the social sciences,'' in \emph{The Science-Music Borderlands}.\hskip 1em plus 0.5em minus 0.4em\relax The MIT Press, 2023, pp. 347--366.

\bibitem{jacoby2020cross}
N.~Jacoby, E.~H. Margulis, M.~Clayton, E.~Hannon, H.~Honing, J.~Iversen, T.~R. Klein, S.~A. Mehr, L.~Pearson, I.~Peretz \emph{et~al.}, ``Cross-cultural work in music cognition: Challenges, insights, and recommendations,'' \emph{Music Perception}, vol.~37, no.~3, pp. 185--195, 2020.

\end{thebibliography}
\fi

\ifdefined\GlobalTapPreprint
  \clearpage
  \onecolumn
  \appendix
  \setcounter{figure}{0}
\setcounter{table}{0}
\renewcommand{\thefigure}{A\arabic{figure}}
\renewcommand{\thetable}{A\arabic{table}}
\renewcommand{\theHfigure}{A\arabic{figure}}
\renewcommand{\theHtable}{A\arabic{table}}

\section{Supplementary Analyses}
\label{app:supplementary}

This appendix contains the supplementary figures, tables, and analyses.
Playhead videos of the patterns in Section~\ref{sec:divergences} are in the
\href{\SupplURL\#audio-demos}{supplementary HTML}, and the underlying audio and
video files are in the \href{\RepoURL}{companion repository}.

\subsection{Per-corpus rating distributions}
\label{app:rating-distributions}

\begin{figure}[H]
  \centering
  \includegraphics[width=0.82\textwidth]{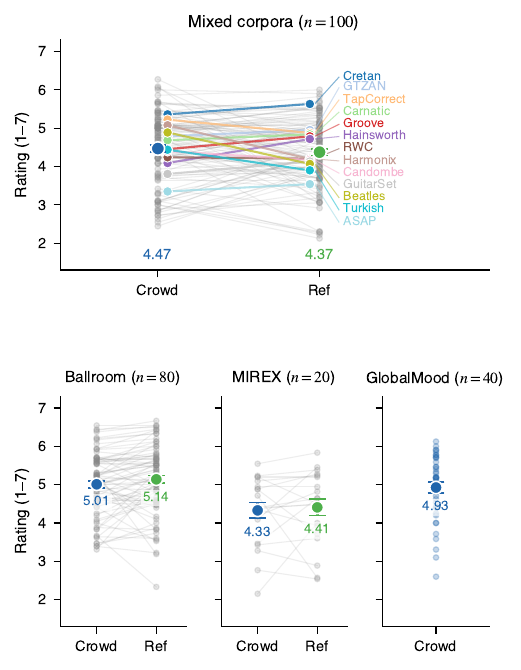}
  \caption{Per-corpus distributions of mean per-excerpt ratings on the 1--7
  scale for Crowd and Reference annotations across the four validation corpus
  groups. The summary statistics from this figure (per-corpus means, $\Delta$,
  $p$-values, and effect sizes) are reported in
  Table~\ref{tab:main}. The mixed-corpora plot also shows the per-source
  Crowd and Reference means for its 13 constituent corpora. Error bars are
  $\pm1$ SE across excerpts.}
  \label{fig:app-rating-distributions}
\end{figure}

\subsection{Per-source results for the mixed corpora}
\label{app:mixed-source}

The 100 mixed-corpora excerpts come from 13 source corpora.
Wilcoxon signed-rank tests compare excerpt-level Crowd and Reference means
within each source, with Benjamini--Hochberg correction across the 13 tests.
None remain significant at $\alpha=.05$ after correction.

\begin{table}[H]
  \centering
  \caption{Crowd and Reference ratings by source corpus in the mixed corpora.
  Crowd and Ref are mean per-excerpt ratings.
  $\Delta=\text{Crowd}-\text{Ref}$. $p_{\mathrm{BH}}$ is corrected across
  the 13 source-corpus tests.}
  \label{tab:app-mixed-source}
  \small
  \setlength{\tabcolsep}{5.5pt}
  \begin{tabular}{lrrrrrrr}
    \toprule
    Source corpus & $n$ & Crowd & Ref & $\Delta$ & $p$ & $p_{\mathrm{BH}}$ & $d_z$ \\
    \midrule
    ASAP          & 8  & 3.35 & 3.54 & $-0.19$ & .742 & .877 & $-0.20$ \\
    Beatles       & 8  & 4.89 & 4.07 & $+0.82$ & .055 & .178 & $+0.83$ \\
    Candombe      & 13 & 4.40 & 4.17 & $+0.23$ & .733 & .877 & $+0.20$ \\
    Carnatic      & 4  & 4.69 & 4.84 & $-0.15$ & .625 & .877 & $-0.30$ \\
    Cretan        & 3  & 5.36 & 5.63 & $-0.27$ & .500 & .812 & $-0.80$ \\
    Groove (MIDI) & 8  & 4.45 & 4.79 & $-0.33$ & .312 & .580 & $-0.44$ \\
    GTZAN         & 8  & 4.79 & 4.95 & $-0.16$ & .844 & .914 & $-0.23$ \\
    GuitarSet     & 8  & 3.81 & 4.13 & $-0.32$ & .031 & .135 & $-0.75$ \\
    Hainsworth    & 8  & 4.09 & 4.72 & $-0.63$ & .008 & .102 & $-1.31$ \\
    Harmonix      & 8  & 5.08 & 4.20 & $+0.88$ & .078 & .203 & $+0.74$ \\
    RWC           & 8  & 4.24 & 4.20 & $+0.04$ & .938 & .938 & $+0.04$ \\
    TapCorrect    & 8  & 5.22 & 4.88 & $+0.34$ & .250 & .542 & $+0.44$ \\
    Turkish       & 8  & 4.44 & 3.90 & $+0.54$ & .016 & .102 & $+1.39$ \\
    \bottomrule
  \end{tabular}
\end{table}

\subsection{Participant demographics and study flow}
\label{app:demographics}

\textbf{Benchmark recruitment and exclusions.}
Across the ten country panels, 2,160 people began a session: 707 completed it,
831 did not pass the tapping performance check, 134 declined a consent step,
and 488 left before finishing. Of the 28,945 recorded trials, 23,493 passed the
recording, marker-analysis, and post-trial checks. The 746 benchmark tappers
are those with at least one such trial (698 who completed and 48 who left
early). Of the 139 people who started the listening experiment, 130 finished
it, and all 6,487 reported ratings come from them.

\begin{table}[H]
  \centering
  \caption{Tapper groups in the benchmark and validation analyses.}
  \label{tab:app-participant-panels}
  \small
  \setlength{\tabcolsep}{5pt}
  \begin{tabular}{llrrrr}
    \toprule
    Excerpt set & Reference source & Excerpts & Tappers &
    Median per excerpt & Trials \\
    \midrule
    Ballroom & Expert beat grids & 80 & 231 & 40 & 3,821 \\
    Mixed corpora & 13 source corpora & 100 & 104 & 36 & 3,555 \\
    MIREX & Lab-tapper consensus (40 participants) & 20 & 102 & 67 & 1,353 \\
    GlobalMood validation & None (absolute-quality check) & 40 & 296 & 110 & 4,403 \\
    GlobalTap benchmark & Crowd-derived & 200 & 746 & 117 & 23,493 \\
    \midrule
    Total & --- & 440 & 1,479 & --- & 36,625 \\
    \bottomrule
  \end{tabular}
\end{table}

The trial totals and median number of tappers per excerpt include only trials
that passed the recording, marker, and post-trial checks.

\begin{table}[H]
  \centering
  \caption{Demographics of the 746 GlobalTap benchmark tappers.}
  \label{tab:app-demographics}
  \small
  \begin{tabular}{p{0.30\textwidth}p{0.60\textwidth}}
    \toprule
    Variable & Distribution \\
    \midrule
    Age (years) & $M=33.1$, $SD=11.1$ \\
    Gender & 49.5\% female, 49.1\% male, 1.4\% non-binary or undisclosed \\
    Formal musical training & Median 1 year (IQR 0--4), with 33\% none and 9\% at least 10 years \\
    Training median by panel & 1--6 years \\
    Gold-MSI (25-item, 1--7) & Overall $M=4.10$, $SD=0.88$, with panel means 3.77--4.56 \\
    \bottomrule
  \end{tabular}
\end{table}

\subsection{Tapper-count convergence}
\label{app:convergence}

Figure~\ref{fig:app-convergence} shows convergence of
participant-normalized tapping KDEs and their extracted peaks. Random
subsamples of tappers were compared with the full consensus they were drawn
from. Density correlation reached $.90$ at 40 tappers and peak F-measure
reached $.90$ at 30 tappers. At 90 tappers, the values were $.985$
and $.971$.

The independent-group comparison in \S2.3 ran two separate groups per excerpt
through KDE, peak extraction, and optimization. Beyond the final-grid
F-measures reported there, the two groups' grids had the same beat rate on
85.0\% of excerpts at $N=20$ and 88.5\% at $N=50$.

\begin{figure}[H]
  \centering
  \includegraphics[width=0.88\textwidth]{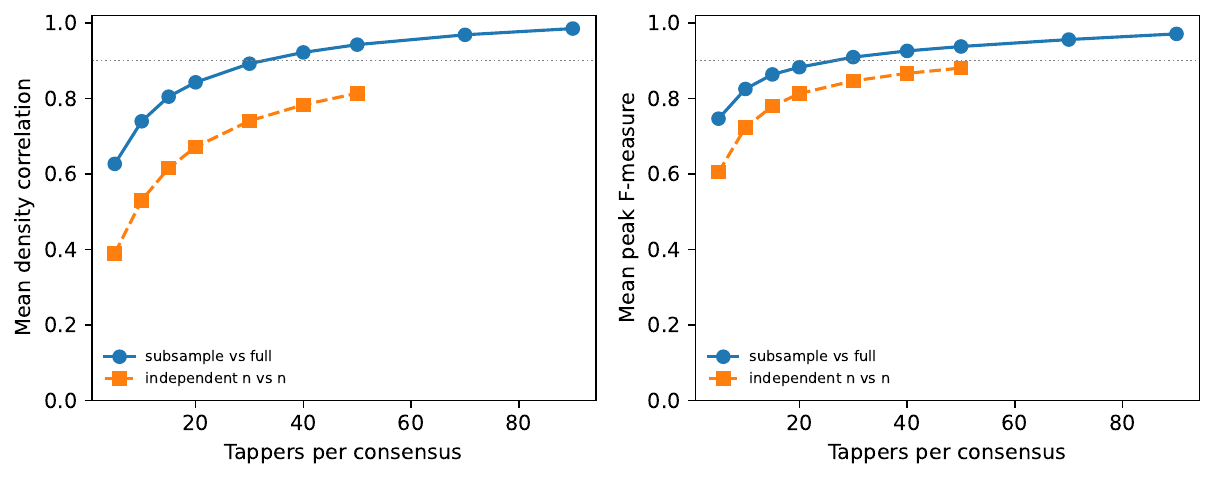}
  \caption{Convergence as a function of tappers per excerpt. Circles compare
  subsamples with the full consensus. Squares compare independent groups.
  The dotted horizontal line marks $.90$.}
  \label{fig:app-convergence}
\end{figure}

\subsection{Exploratory country-panel analyses}
\label{app:country-panels}

For each excerpt, each trial's tapping period was compared with a reference
period, the median of the other nine panels' median periods, and classified as
double rate (ratio .35--.65), matched rate (.80--1.25), half rate
(1.75--2.25), or other/unclear. The period analysis used 23,492 trials (one of
the 23,493 had too few taps for a period estimate) and the phase analysis the
11,450 matched-rate trials. Panels were compared by subtracting each excerpt's
mean from every trial's outcome, averaging the remainder within each
participant, and permuting panel labels across participants (9,999
permutations).

For the own-country test in the five focal countries, ISTC grows with the
number of tappers, so for each excerpt and panel it was averaged over 500
random samples of seven tappers, the most available for every excerpt and
panel. The regression had a separate baseline for each excerpt and each panel,
with standard errors clustered by excerpt.

\subsection{Metrical-preference analysis}
\label{app:metrical}

This analysis includes all 180 excerpts with independent expert beat
references. Excerpts are grouped by the Crowd annotation's period relative to
the reference. Rating differences are excerpt-level Crowd minus Reference means. An
excerpt is classified as matched, double, or half rate when its
crowd-to-reference period ratio falls within $\pm0.15$ octaves (about
$\pm11$\%) of $1\times$, $0.5\times$, or $2\times$, respectively. All remaining
excerpts are grouped as other ratios.

On matched-rate excerpts, an excerpt's phase offset is the median signed
distance from each Crowd beat to the nearest reference beat within $150$\,ms,
over the rendered $10$--$25$\,s window, and it counts as consistent at
$20$\,ms or more. On $8$ excerpts no Crowd beat came within $150$\,ms of a
reference beat: the Crowd beats fell a third to half a beat from the
reference's. These also count as consistent offsets.

\begin{table}[H]
  \centering
  \caption{Rating differences by Crowd-to-Reference period relation.}
  \label{tab:app-metrical}
  \small
  \begin{tabular}{lrrr}
    \toprule
    Period relation & $n$ & Crowd higher & Median $\Delta$ \\
    \midrule
    Matched rate & 145 & 51 & $-.143$ \\
    Crowd half rate & 9 & 9 & $+.748$ \\
    Crowd double rate & 14 & 7 & $-.238$ \\
    Other ratio & 12 & 7 & $+.393$ \\
    \bottomrule
  \end{tabular}
\end{table}

\subsection{Beat metrics beyond F-measure}
\label{app:beat-metrics}

Table~\ref{tab:app-beat-metrics} reports standard \texttt{mir\_eval} metrics
for the 198 excerpts evaluable for both trackers.

\begin{table}[H]
  \centering
  \caption{Beat-tracking agreement with the GlobalTap consensus.}
  \label{tab:app-beat-metrics}
  \small
  \begin{tabular}{lrr}
    \toprule
    Metric & madmom & Beat This! \\
    \midrule
    CMLc & .797 & .621 \\
    CMLt & .807 & .636 \\
    AMLc & .947 & .908 \\
    AMLt & .956 & .935 \\
    Information gain & .669 & .643 \\
    \bottomrule
  \end{tabular}
\end{table}

Paired excerpt-level tests favored madmom for all five metrics, with FDR
correction across all ten \texttt{mir\_eval} metrics (the other five are in
the repository). CMLc, CMLt, and information gain reached $q<.001$, AMLc
$q=.011$, and AMLt $q=.024$.
Pooled (pre-optimization) ISTC was associated with agreement for madmom
($\rho=.39$, $p<.0001$) and Beat This! ($\rho=.27$, $p=.0001$).
Pipeline-output CV was also associated with agreement for Beat This!
($\rho=-.24$, $p=.0006$) and madmom ($\rho=-.18$, $p=.009$).
The madmom correlations use all 200 excerpts and the Beat This! correlations
the 198 it could evaluate.

\subsection{Grid optimizer settings}
\label{app:impl-clarification}
Table~\ref{tab:app-optimizer} lists the settings of the grid optimizer in
\S2.3. $T$ is the grid period. The companion repository describes every setting
with the code (\texttt{supplementary\_analyses/METHOD\_PARAMETERS.md}).

\begin{table}[H]
  \centering
  \caption{Grid optimizer settings.}
  \label{tab:app-optimizer}
  \small
  \begin{tabular}{p{0.16\textwidth}p{0.76\textwidth}}
    \toprule
    Step & Setting \\
    \midrule
    Search & Period $0.28$--$1.25$\,s ($48$--$214$\,BPM) in $120$ steps, phase in $240$ steps per period. $C$ is the proportion of peaks within $0.12T+20$\,ms of a grid beat. The density has an $80$\,ms Gaussian at each peak, scaled to a maximum of 1 \\
    Refinement & Least squares on peaks within $0.15T$ of the grid, then the phase re-searched within $\pm0.1T$, keeping only phases in $[0, T)$ \\
    Snapping & A beat with a peak within $0.12T$ moves 45\% of the way toward it \\
    Rate doubling & When more than 55\% of the midpoints between beats have a peak within $0.18T$, the mean density at the midpoints is over 40\% of that at the beats, and the doubled rate is at most $214$\,BPM. The beats of the doubled grid are then nudged toward even spacing \\
    Gap filling & Beats inserted in gaps over $1.4T$ and out to the window edges, at the median beat interval \\
    Split peaks & Tried when the largest gap between the sorted inter-peak intervals shorter than $0.9T$ exceeds $0.10T$. The grids are compared by $2C'+(1-\mathrm{CV})+0.5\,\bar{D}$, where $C'$ is the proportion of peaks within $0.12T+30$\,ms of a beat and CV is the coefficient of variation of the beat intervals \\
    \bottomrule
  \end{tabular}
\end{table}

\fi

\end{document}